\documentclass[%
 aps,
 amsmath,amssymb,
reprint,%
]{revtex4-1}

\usepackage{graphicx}
\usepackage{dcolumn}
\usepackage{bm}
\usepackage{blkarray}
\usepackage[usenames,dvipsnames]{color}
\usepackage{multirow}
\usepackage[hyperfootnotes=false]{hyperref}
\usepackage[hyperfootnotes=false]{hyperref}
\hypersetup{
 colorlinks,
 citecolor=Violet,
 linkcolor=Red,
 filecolor= green,
 urlcolor=Blue}
\begin{document}

\title{First-principles study of the electrical and lattice thermal transport in monolayer and bilayer graphene}

\author{Ransell D'Souza}%
\email{ransell.d@gmail.com; ransell.dsouza@bose.res.in}

\author{Sugata Mukherjee}%
\email{Corresponding author: sugata@bose.res.in; sugatamukh@gmail.com}

\affiliation{Department of Condensed Matter Physics \& Materials Science, S.N. Bose National Centre for Basic Sciences, Block JD, Sector III, 
Salt Lake, Kolkata 700098, India}

\begin{abstract}
We report the transport properties of monolayer and bilayer graphene from first principles calculations and Boltzmann transport theory (BTE). 
Our resistivity studies on monolayer graphene show  Bloch-Gr${\rm \ddot{u}}$neisen behavior in a certain range of chemical potentials. 
By substituting boron nitride in place of a carbon dimer of graphene, we predict a twofold increase 
in the Seebeck coefficient. A similar  increase in the Seebeck coefficient for bilayer graphene under the influence of a small electric 
field $\sim 0.3$ eV has been observed in our calculations. Graphene with impurities shows a systematic decrease
of electrical conductivity and mobility. We have also calculated the lattice thermal conductivities
of monolayer graphene and bilayer graphene using phonon BTE which show excellent agreement with experimental data available in the temperature range 300-700\ K.

\end{abstract}

\maketitle

\section{Introduction}
Transport properties in monolayer graphene (MLG) and bilayer graphene (BLG) have drawn a great deal of theoretical and experimental attention in recent years due to 
interest in both fundamental physics and potential electronic applications.
An interesting transport property is the thermoelectric effect (TE) in which a temperature difference $\Delta T$ across a conductor creates an electric potential 
$\Delta V = S\Delta T$ where $S$ is the thermoelectric power, also called Seebeck coefficient, and characterizes the thermoelectric sensitivity of the conductor. 
Thus a material with a high value of $S$ is a necessary condition for it to be used in thermoelectric applications. The performance of the thermoelectric 
material in such TE devices is determined by a dimensionless parameter $ZT = \frac{S^2\sigma}{\kappa}T$ , where $\sigma$ is the electrical conductivity, 
$\kappa = \kappa_e + \kappa_p$ is the thermal conductivity including both electron and phonon contributions, and $T$ is the temperature. 
Hence such materials should have a high power factor $S^2\sigma$ and a low $\kappa$ \cite{rowe06, tritt01, mahan98}.

Extensive experiments have been carried out by Kim $et\ al$ \cite{kim, kim2} on transport properties in graphene.
For ultraclean suspended graphene, Bolotin $et\ al.$\cite{kim2} have shown that resistivity is strongly temperature dependent. 
At low temperatures ($T\sim 5K$) the electron transport is near-ballistic having a mobility of $\sim$ 170000 ${\rm cm^2}/{\rm Vs}$. 
For large carrier densities, resistivity increases with $T$ and is linear above 50K. 
This suggests that the carrier scattering is from acoustic phonons. They also reported the highest measured value of mobility ($\sim$ 120000 ${\rm cm^2}/{\rm Vs}$) 
at $T$=240K. 
Using a micro-fabricated heating system, electrical conductivity and Seebeck coefficients were measured for the first time by 
Zuev $et\ al$ \cite{kim}.  
The gate-dependent and hence carrier-density-dependent conductance and Seebeck coefficient showed good agreement with the semiclassical Mott relation. 
It was shown that the sign of Seebeck coefficients changes as the majority carrier density shifts from the electron to that of the hole. 
The gate-dependent Seebeck coefficient and Nernst signal on graphene in a magnetic field was studied by Ong $et\ al.$ \cite{ong09}.
Quantities closely related to $\sigma$ are mobility ($\mu_{FE}$) and resistivity ($\rho$). Since a field effect transistor can measure the mobility, 
we use $\mu_{FE}$ to define mobility so as to not confuse it with chemical potential ($\mu$).  Values of $\mu_{FE}$ as high as 20000 ${\rm cm}^2/{\rm V\ s}$, 
have been reported for MLG at low temperatures $T$ \cite{novoselov05, zhang05, schedin07, tan07, chen08}. It is well known that the resistivity $\rho$ strongly 
depends on the electron-phonon (e-ph) scattering in the material and varies linearly with temperature, $\rho \sim T$, at temperatures comparable to or higher than the 
Debye temperature 
$\Theta_D$ , reflecting a classical behavior of the phonons. The bosonic nature of the phonons becomes evident below $\Theta_D$ resulting in a rapid decrease of 
resistivity which for a normal conductor $\rho \sim T^5$, known as the Bloch-Gr${\rm \ddot{u}}$neisen (BG) regime. This defines a new characteristic temperature 
scale for the low-density e-ph scattering, the BG temperature $\Theta_{BG} = \frac{2v_{ph} k_F}{k_B} < \Theta_D$, where $v_{ph}$ and $k_F$ are the phonon group
velocity and Fermi momentum, respectively. It was theoretically predicted \cite{sarma08} 
that in a two-dimensional systems such as graphene, there is a smooth transition in the resistivity from a linear dependence $\rho \sim T$ to $\rho \sim T^4$. 
This two-dimensional (2D) BG was observed experimentally by Efetov and Kim \cite{efetov10} in graphene.

Temperature dependence of electron transport in graphene and its bilayer have also been studied by Morozov $et\  al.$ \cite{morozov08}. 
They have shown that if extrinsic disorders are eliminated, mobilities higher than $2\times10^5 {\rm cm^2/Vs}$ are attainable.
Ponomarenko $et\ al.$ \cite{ponomarenko09} studied the scattering mechanism  on graphene on various substrates and found no significant changes in carrier 
mobility on different substrates questioning the largely believed assumption 
that the dominant source of scattering in graphene is the charge impurities in the substrate.

The experimental works carried out by various groups such as Kim $et\ al$ \cite{kim}, Geim $et\ al$\cite{ponomarenko09}, Ong $et\ al$ \cite{ong09}, 
Nam $et\ al$ \cite{nam10} and Wang $et\ al$ \cite{wang11} on gate-dependent electron transport properties of MLG and BLG have motivated us to carry out computational 
studies based on first-principles calculations of MLG-doped hexagonal boron nitride (h-BN) and pure BLG under the influence of an electric field, since there is a direct 
correspondence between gate voltage and chemical potential. 
It is observed that $S$ is strongly dependent on the amount of doping in the case of MLG. A marked increase of $S$ is observed when an electric field is 
applied perpendicularly to the plane of the bilayer graphene sheets.

A further motivation to study the effects of impurity scattering on
graphene sheets was provided by a recent paper by Ghahari $et\ al$ \cite{kim2016}, which reported an
enhanced Seebeck coefficient in graphene due to scattering.
We have used a tight-binding model (TBM) for graphene with impurities and solved the Boltzmann transport equations for electrons to study the effect of impurities
on electrical conductance.
Our study used a large $k$ mesh to capture correctly the enhancement of the 
Seebeck coefficient and a constant decrease in electrical conductivity due to doping and decrease in mobility by an order of magnitude as observed 
experimentally \cite{kim2016,chen08,tan07}.

Finally, we have reexamined the lattice thermal conductivity in MLG and BLG using the lattice Boltzmann transport method \citep{ShengBTE} employing the phonon 
bandstructure calculated from first-principles DFT \cite{giannozzi09}. Our calculated phonon dispersion agrees with previous results 
\cite{yan2008, nika2009, kong2009, yao2000, maultzsch2004}, whereas the lattice thermal conductivity shows quantitative agreement with recent experimental data 
\cite{hongyang2014}.

The purpose of this paper is to illustrate that all essential thermoelectric parameters of MLG and BLG, both for electrical and lattice thermal transport, can be accurately
calculated using first-principles electronic and phonon band structure together with the Boltzmann transport theory. Our results are in excellent agreement with
experimental data.

\section{Method of Calculation}

\subsection{Density functional theory based electronic structure calculation}

All electronic structure calculations were carried out using the density functional theory (DFT) based plane-wave
method as implemented in the Quantum Espresso code \cite{giannozzi09}.
A hexagonal unit cell has been used in all these calculations. For the exchange-correlation potential, we use the generalized gradient approximation (GGA) \cite{pbe96} and the ultrasoft pseudopotential \cite{vanderbilt90} to describe the core electrons.
Monkhorst–Pack $k$-point grids \cite{mp76} of $120 \times 120 \times 1$ and $70 \times 70 \times 4$  were implemented for monolayer and bilayer graphene 
in the self-consistent calculations with a 40Ry kinetic energy cutoff and a 160Ry charge density energy cutoff, respectively. The Kohn-Sham equations are
solved self-consistently to achieve an accuracy of $10^{-9}$ Ry in the total energy.
Plane-wave methods incorporate periodicity and hence in order to avoid interaction between sheets we have a vacuum spacing of 22{\AA} along the $z$ direction.
The van der Waals interaction was included for multilayered systems.

\subsection{Boltzmann transport theory for band electrons}

The semiclassical Boltzmann transport theory (BTE) applied to band electrons, as implemented in the Boltztrap code \cite{boltztrap1}, was used to calculate 
the transport properties. The BTE allows us to calculate the thermoelectric parameters along two orthogonal principal axes in the plane of the two-dimensional
graphene layers. Thus the calculated thermoelectric parameters are taken as average over those along the principal directions.
The energies and their corresponding k-points were taken from the electronic structure calculations to deduce various transport parameters.
For example, the group velocity of the electrons $v_{\alpha}$ is given by,
\begin{eqnarray}\label{v}
v_{\alpha}(i,\textbf{k})=\frac{1}{\hbar}\frac{\partial \varepsilon_{i,k}}{\partial k_{\alpha}}
\end{eqnarray}
In the relaxation-time approximation electrical conductivity is then expressed as,
\begin{eqnarray}\label{sig}
{\sigma_{\alpha\beta}(T,\mu) \over \tau} &=& {1 \over V} \int e^2 v_{\alpha}(i,{\bf k})\, v_{\beta}(i,{\bf k}) [{-\partial f_\mu(T,\epsilon) \over \partial \epsilon}] d\epsilon
\end{eqnarray}
Detailed studies \cite{wwschulz, pballen2} have shown that to a good approximation the relaxation time $\tau$ is isotropic.

The Seebeck coefficient is then calculated using,
\begin{eqnarray}\label{S}
S_{\alpha \beta}(T,\mu) = \frac{1}{eT}\frac{\int v_{\alpha}(i,{\bf k}) v_{\beta}(i,{\bf k})(\epsilon - \mu) [{-\partial f_\mu(T,\epsilon) \over \partial \epsilon}]d\epsilon}
{\int v_{\alpha}(i,{\bf k}) v_{\beta}(i,{\bf k}) [{-\partial f_\mu(T,\epsilon) \over \partial \epsilon}]d\epsilon}
\end{eqnarray}
where, $f_{\mu} = \frac{1}{1+ e^{\beta (\epsilon-\mu)}}$ is the Fermi-Dirac distribution, $\mu$ is the chemical potential and $k_B$ is the Boltzmann constant.
MLG and BLG are known to be semi-metallic where the transport occurs only near the Fermi level. We can therefore use the Sommerfeld expansion of Eqs. \ref{sig} and 
\ref{S} to obtain,
\begin{eqnarray}\label{Smott}
S = -\frac{\pi^{2} k_B^2T}{3e} {d \over dE}\, [ln\, \sigma(E)]{\Bigr|}_{E=E_F}
\end{eqnarray}
Eq. \ref{Smott} is known as the Mott formula \cite{mott}.

\subsection{Transport properties from tight-binding method}
In order to study the effect of dilute impurities in graphene on its electrical conductivity, we have used a tight-binding (TB) model of graphene
using the method described in \cite{saito98} for MLG and \cite{mccann13} for BLG. Thermoelectrical parameters are then calculated from
the Boltzmann transport equations described above. The TB model allows us to calculate using a very large unit cell containing hundred atoms or more with
$k$ pont grids of $120\times120\times1$ for MLG and $70\times70\times4$ for BLG, not accessible by the DFT-based plane wave methods.

\subsection{Lattice thermal conductivity from BTE}

We have used the phonon Boltzmann transport method to solve the Boltzmann transport equations for phonons starting from a set of interatomic force constants (IFCs) obtained 
from the phonon dispersion obtained using {\it ab initio} calculations \cite{giannozzi09}, as implemented in the ShengBTE code \cite{ShengBTE}.
Lattice thermal conductivity calculations require the harmonic second-order interatomic force constants (IFCs) as well as the anharmonic  third-order IFCs. 
The harmonic IFCs were calculated using density functional perturbation theory. To attain the anharmonic IFCs, we use a real-space supercell approach. 
The phonon Boltzmann transport calculates the converged set of phonon scattering rates and uses them to obtain the lattice thermal conductivity $\kappa_{L}$ using the expression, 
{\small \begin{eqnarray}\label{kl}
\kappa_L^{\alpha \beta}=\frac{1}{k_BT^2\Omega N}\sum_{\lambda}f_0(f_0+1)(\hbar \omega_\lambda)^2v_{\lambda}^{\alpha} \tau_{\lambda}^0 (v_\lambda^{\beta}+\Delta_\lambda^{\beta})
\end{eqnarray}}
where, $k_B$ is the Boltzmann constant, $\Omega$ is the volume of the unit cell, $N$ is total number of $q$ points in the Brillouin zone sampling, $\tau_{\lambda}^0$ is the relaxation time of mode $\lambda$ which is obtained from
perturbation theory and used within the relaxation time approximation (RTA) and $f_0$ is the Bose-Einstein statistic. $\omega_\lambda$ and $v_\lambda$ are the angular frequency and group velocity of the phonon mode $\lambda$, where $\lambda$ comprises of a phonon branch index and wave vector. $\Delta_{\lambda}$ is the measure of how much the population  of a specific phonon mode (and hence associated heat current) deviates from the RTA prediction.
Detailed information on the work-flow can be found here \cite{ShengBTE}.

For both cases, MLG and BLG, a $4\times 4 \times 1$ supercell along with a $5\times 5 \times 2$ Monkhorst–Pack $k$ mesh was chosen to evaluate 
the forces. All calculations involved in the anharmonic third-order IFCs accounted for interactions up to the third-nearest neighbors. 
The phonon calculations were performed using an ultrasoft pseudopotential \cite{rrkj}, which reproduces correctly the phonon dispersion of 
both MLG and BLG.
In the present study the van der Waals interaction was included for BLG as prescribed by Grimme \cite{grimme2}.

\section{Results and Discussion}
\subsection{Electrical conductivity and Seebeck coefficient of MLG}
\begin{figure}[!htbp]
\includegraphics[scale=0.3]{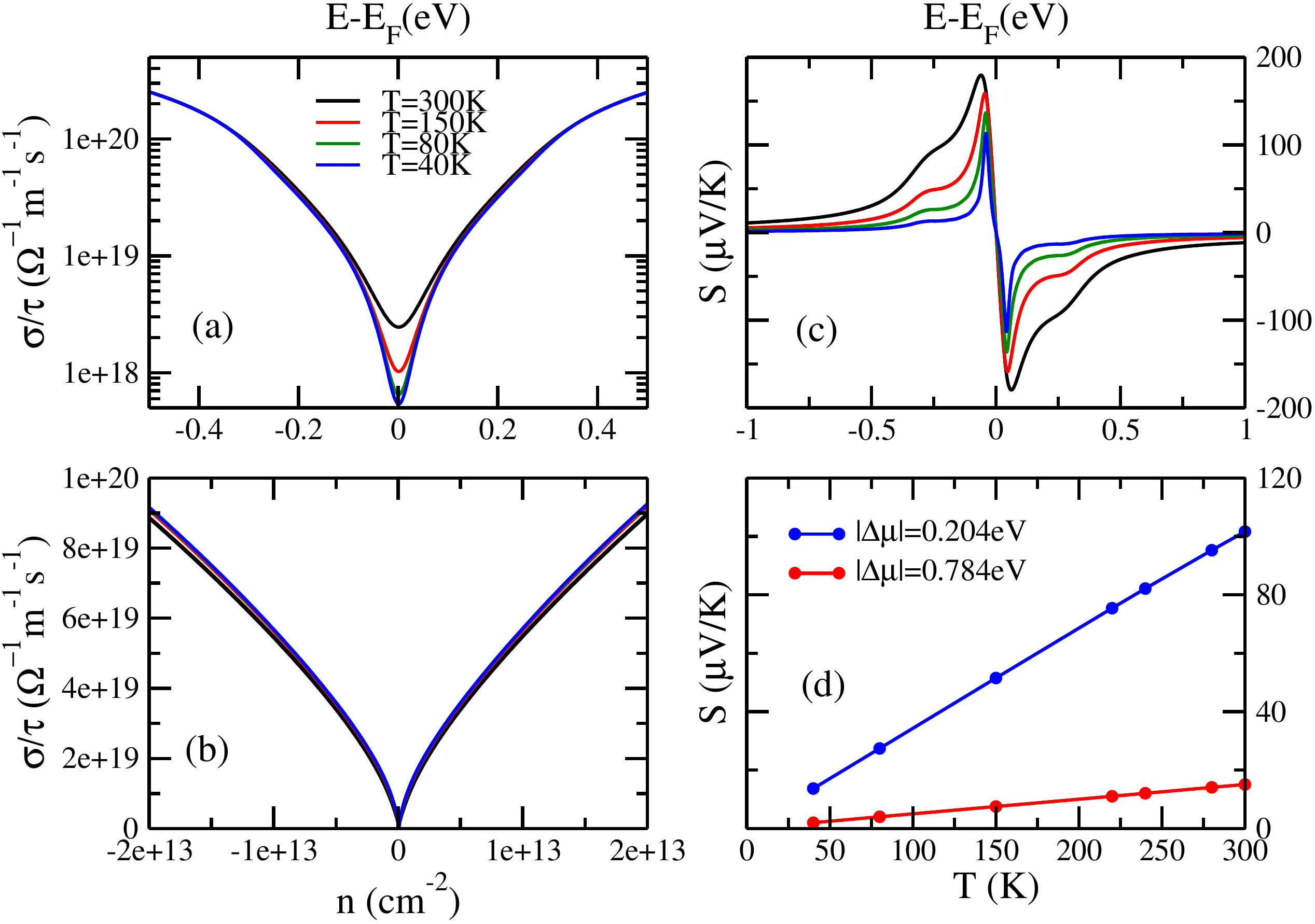}
\caption{\label{sigma} (a) Scaled electrical conductivity ($\sigma$/$\tau$) as a function of energy at different temperatures, (b) $\sigma$/$\tau$ as a function of 
charge carriers $n$ at different $T$, (c) Seebeck coefficient($S$) as a function of energy at different $T$ and (d) $S$ as a function of $T$ at two values of $\mu$.}
\end{figure}
The calculated lattice constants for both MLG and BLG were 2.47 \AA\ using GGA and 2.46 \AA\ using LDA, respectively. The interlayer separation for BLG was found
to be 3.32 \AA.

As the precise numerical value of the electron relaxation time $\tau$ for graphene is not known and can only be estimated 
from experiments \cite{tan07}, we show in Figs. 1(a,b) the scaled electrical conductivity $\sigma$/$\tau$, calculated from the Bolzmann transport theory (Eq. {\ref{sig}}), 
as a function of energy and carrier concentration in the temperature range 40K$-300$K, respectively.  
It is observed that close to the Fermi energy $\epsilon_F$, $\sigma$ increases with increasing temperature but is almost independent of $T$ away from Fermi energy.
This behavior has been reported by Morozov $et\ al$. \cite{morozov08}.
In Fig. \ref{sigma}(b) we observe $\sigma$ of MLG to be proportional to $\sqrt{n}$ where $n$ is the charge carrier density.

This $\sqrt{n}$ behavior can be explained using a tight-binding model with an energy expansion  around the $k$ point to get $\epsilon_F=\hbar v_F k_F$, 
where $\hbar$ is the reduced Planck constant, $v_F$ is the Fermi velocity and $k_F$ is the Fermi wavenumber. For a 2-dimensional sample, 
\begin{eqnarray}
n = \frac{k_F^2}{\pi} = \frac{\epsilon_F^2}{\pi(\hbar v_F)^2}
\end{eqnarray}
Therefore the MLG's density of states $D(\epsilon_F)$ is given by,
\begin{eqnarray}
D(\epsilon_F) = \frac{dn}{d\epsilon}=\frac{2\sqrt{n}}{\hbar v_F \sqrt{\pi}}
\end{eqnarray}
Substituting this in the Einstein relation $\sigma \propto D(\epsilon_F)$, we obtain $\sigma \propto \sqrt{n}$. This behavior has been experimentally confirmed for 
pristine graphene \cite{kim2}.

The Seebeck coefficient was calculated using the Mott relation Eq. \ref{Smott} numerically and is plotted in Fig. \ref{sigma}(c) as a function of energy. Our calculated 
form of $S$ is in very good agreement with the experimental results reported by Zuev $et\ al$ \cite{kim}.
Using a back-gated field effect transistor, the Fermi energy can be tuned by adjusting the gate voltage. Zuev $et\  al$ \cite{kim} have used this method for a mesoscopic 
graphene sample using the formula $S=-\frac{\pi^{2} k_B^2T}{3e}\frac{1}{\sigma}\frac{d\sigma}{dV_g}\frac{dV_g}{dE}|_{E=E_f}$. It can be easily seen that this is 
identical to Eq. \ref{Smott}. Therefore there is a direct correspondence between the gate voltage and the chemical potential.
In Fig. \ref{sigma}(d) we plot $S$ against $T$ for two different values of chemical potential. 
The linear dependence of $S$ on $T$, which has also been reported by experimental measurements by Zuev $et\ al.$ \cite{kim}, suggests that the mechanism of thermoelectric 
transport is diffusive \cite{mahan02}.

\subsection{Resistivity and fit to the Bloch-Gr\"uneisen form}

\begin{figure}[!htbp]
\centering \includegraphics[scale=0.3]{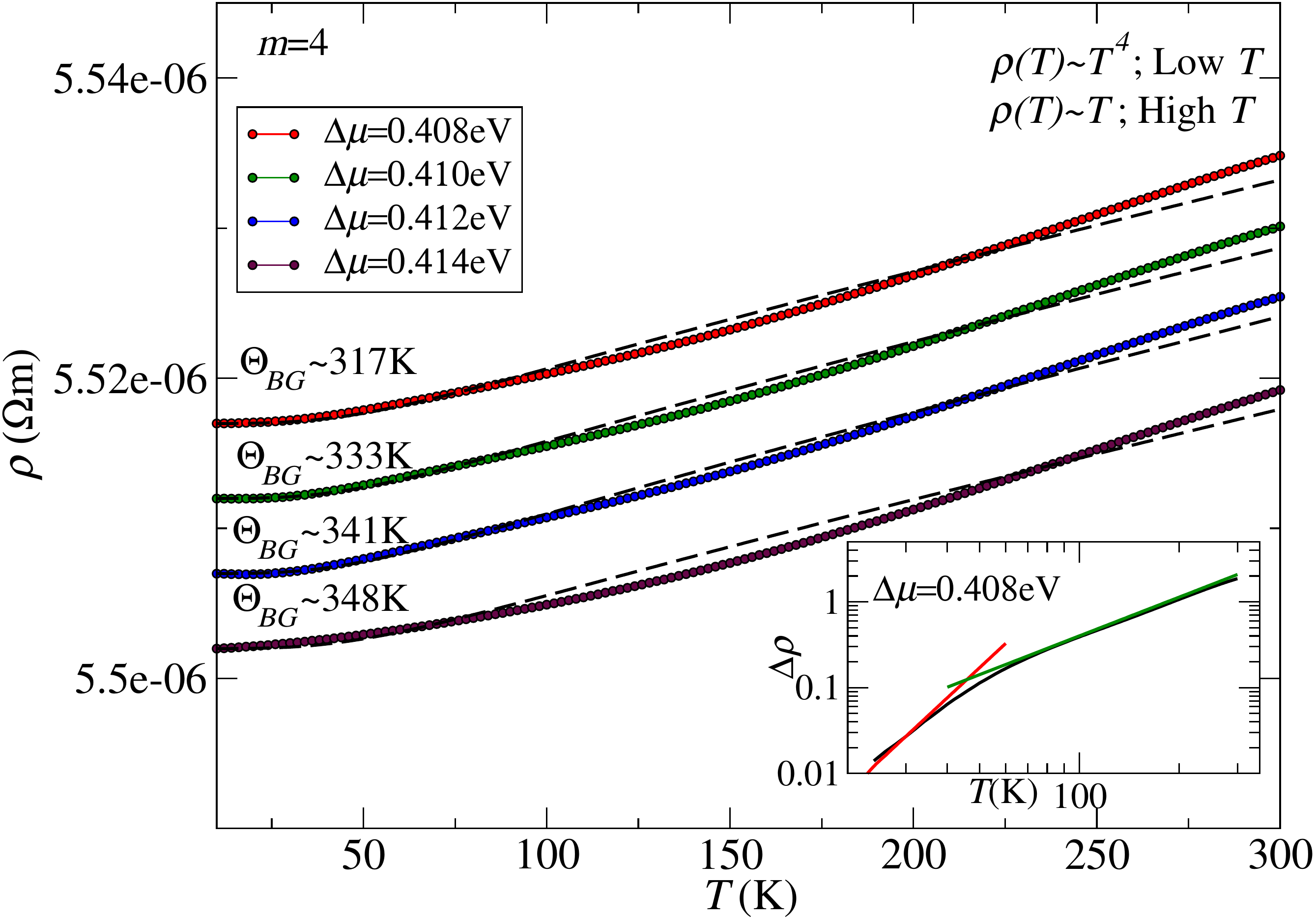}
\caption{\label{rho} Calculated resistivity $\rho$ as a function of $T$ for MLG at different chemical potential. Dashed black lines are the best fit of the 
Bloch-Gr${\rm \ddot{u}}$neisen formula. Points refer to our calculated values of $\rho$ from the Boltzmann transport equations, assuming $\tau = 1\times 10^{-14}\ $s. 
Inset: $\Delta \rho$ plotted against $T$ in logarithmic scale to highlight 
the $T^4$ and $T$ features. The red and green lines are equations of $\propto T^4$ and $\propto T$. The curves referring to different $\Delta \mu$ are shifted slightly 
along the $y$ axis for clarity.}
\end{figure}

In Fig. \ref{rho} we plot  $\rho$ as a function of $T$ calculated from the Boltzmann transport equations for electron transport. 
The electron relaxation time $\tau$ of graphene depends on the degree of doping and carrier concentration as indicated by experimental
measurements \cite{tan07} and can vary in the range 10$\ $fs to 1$\ $ps. For simplicity we have assumed
$\tau \sim 1\times 10^{-14}\ $s; however, our temperature dependent behaviour of $\rho$
should not depend on the choice of $\tau$.  
The behaviour of $\rho$ at certain values of $\mu$ having the Bloch-Gr${\rm \ddot{u}}$neisen form has been observed experimentally \cite{efetov10, kim2}. 
In order to understand the behaviour, we fit our calculated values of $\rho$ to the Bloch-Gr${\rm \ddot{u}}$neisen formula,
{\footnotesize	
\begin {eqnarray}
\rho(T)=\rho(0) + A\Big(\frac{T}{\Theta_{BG}}\Big)^m\int^{\frac{\Theta_{BG}}{T}}_0 \frac{x^m}{(e^x-1)(1-e^{-x})}dx
\end{eqnarray}
}
\\ with $m$ and $T$ as fitting parameters. 
Our best fit resulted in $m=4$ and $\Theta_{BG}$ as shown in Fig. \ref{rho}. 
This suggests that in the low-$T$ regime, resistivity scales as $T^4$ and smoothly scales to a linear $T$ behavior at higher $T$ regimes. The $T^4$ behavior of 
resistivity reflects the two-dimensional nature of electrons in graphene.
At high temperatures the quantization of lattice waves is irrelevant; therefore the scattering is proportional to the square of the amplitude of the fluctuations about 
their equilibrium position that is proportional to $\sqrt{T}$, which leads to the linear behavior of resistivity at higher temperatures.
The reduction from the $T^5$ behavior of the resistivity for a typical bulk metal to the $T^4$ one, as seen in graphene, is due to reduced electron Fermi surface of graphene, 
smaller than 
the size of its phonon Brillouin zone. Therefore, only a small fraction of acoustic phonons will scatter off the electrons. An excellent illustration and explanation 
of the high, low and Bloch-Gr${\rm \ddot{u}}$neisen temperature regimes, by taking into account different sizes of the Fermi surface of graphene and a conventional metal at 
different temperatures, was given by Fuhrer \cite{fuhrer10}.

As mentioned previously, $\Theta_{BG}$ is defined as $\Theta_{BG}=\frac{2v_{ph}k_F}{k_B}$, where $v_{ph}$ is the phonon velocity. However when expressed as a function 
of carrier concentration, it can be easily shown \cite{sdsarma} to have the form $\Theta_{BG}=A_0 \sqrt{n}\,$K, with carrier concentration measured in $10^{12}$cm$^{-2}$. 
In our calculation, when fitted to the above form, we obtain $A_0$=45.5, which is close to the earlier estimated value of 54 \cite{sdsarma}.

In the inset of Fig. \ref{rho} we plot $\Delta \rho$ ($ = \rho(T) - \rho(0)$) against temperature. 
This parameter will now show us the increase in resistivity with respect to temperature and when plotted in logarithmic scale, we can see that the increase in resistivity 
in the lower regime has a $\sim T^4$ feature and $\sim T$ feature at higher temperatures. The red and green lines in the inset are $\propto T^4$ and $\propto T$ 
equations where the constant of proportionality was found by a best  fit method.
$\Theta_{BG}$ is in the same order of magnitude range as reported by \cite{efetov10}. 
We must note that this behavior of $\rho$ is found only for a certain range of chemical potential. 
If we compare our Seebeck coefficient result to that of the experimental results by Zuev $et\ al$ \cite{kim} we find that as we vary our chemical potential from 
1 eV to -1 eV, the same behaviour is found when they vary their gate voltage from 40V to -40V. This implies that a small change in the chemical potential is equivalent 
to a large change in gate voltage. This is the reason  why we choose only a small range of chemical potentials to demonstrate the Bloch-Gr${\rm \ddot{u}}$neisen 
behavior as observed experimentally. If one looks at the study by Kim $et\ al$ \cite{kim2}, the linear part of resistivity (higher $T$ regime) increases 
quickly with an increase of gate voltage of only 1eV. As observed experimentally \cite{efetov10}, we found that the slope of the $\rho$ vs $T$ curves in the linear-$T$ 
regime increases with $\mu$. 
The present study of the temperature dependent resistivity, which has been theoretically predicted \cite{sarma08} and experimental observed \cite{efetov10}, 
is here done using the Boltzmann transport theory applied to band electrons. 

\subsection{Enchancement of Seebeck coefficient upon doping and in presence of an electric field}
As mentioned previously, the performance of a thermoelectric material is measured by a dimensionless parameter, the figure of merit($ZT=\frac{S^2\sigma}{\kappa}$). Hence a technique to increase the Seebeck coefficient and simultaneously decrease the thermal conductivity is highly desired. Very recent studies \cite{byoung15, elaheh15} have shown that doping of and impurities in graphene sheets will decrease the thermal conductivity. Pop $et\ al.$ \cite{pop2012} have mentioned that any surplus residue from sample fabrication or any form of disorder will reduce the thermal conductivity further. 
There have been many experimental \cite{kan,ding,dutta,pruneda,bhowmick,liu} and theoretical \cite{grossman12,cnrrao13,RDSM} reports to show the formation of a band gap by doping graphene with boron nitride. Similarly there are reports to show the formation of band gaps in bilayer graphene \cite{hongki07,castro07,kin09} when under the influence of an electric field. 
We have successfully used the Boltzmann transport equation for electron transport to calculate the thermoelectric parameter of C$_x$(BN)$_{1-x}$ \cite{RDSM} and 
graphene/h-BN/graphene heterostructures \citep{RDSM16}.
Therefore in this section, using the Boltzmann transport equations, we study the behavior of these transport parameters focusing mainly on enhancing the Seebeck coefficient. 

\begin{figure}[!htbp]
\centering \includegraphics[scale=0.3]{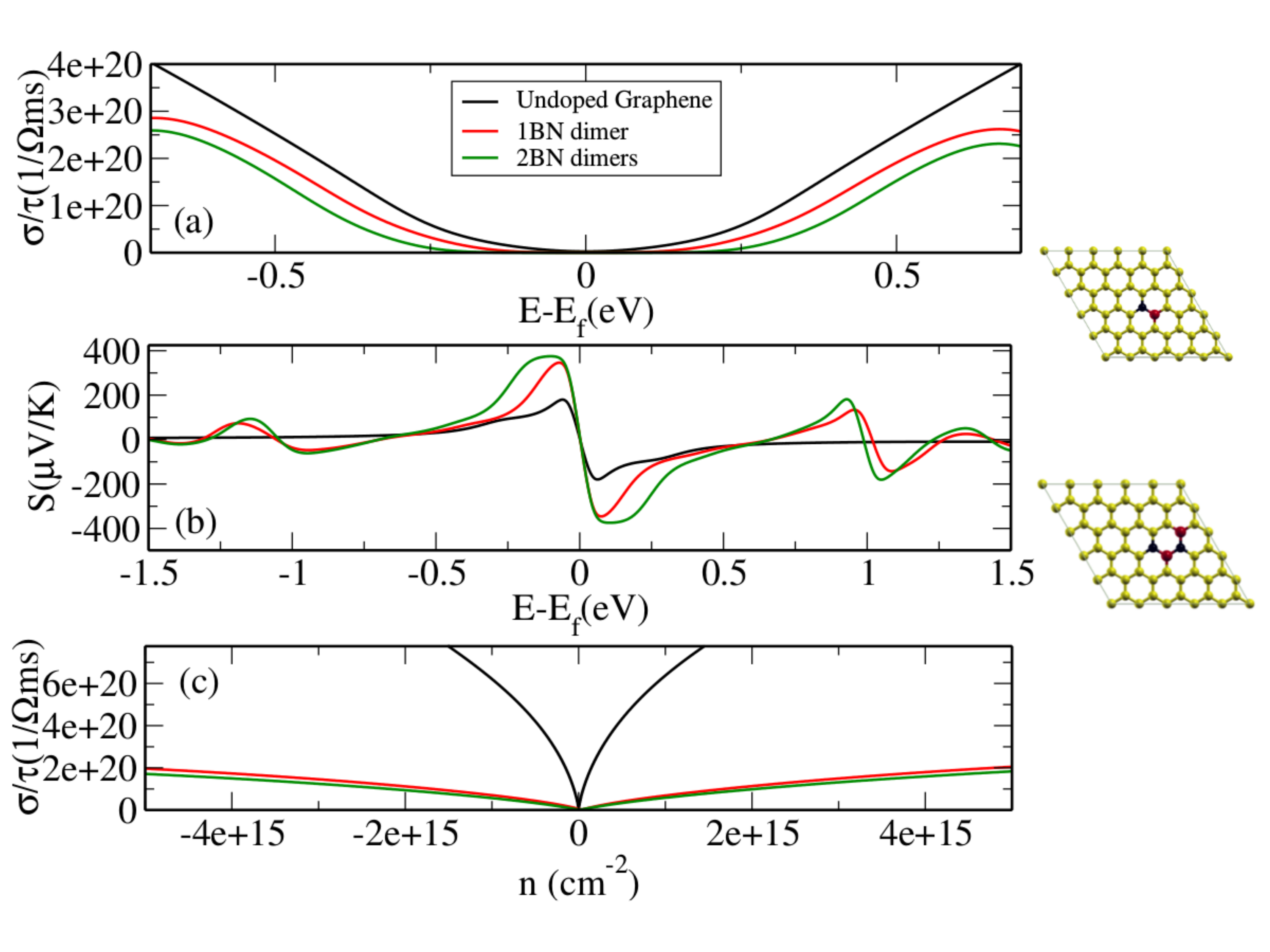}
\caption{\label{dg} (a) $\sigma/\tau$ plotted as a function of energy. (b) Seebeck coefficient plotted as a function of energy. (c) $\sigma/\tau$ plotted as a function 
of $n$. The graphene supercell containing one and two BN dimer are shown on the right.}
\end{figure}

In Fig. \ref{dg} we show the calculated electrical conductivity ($\sigma$) and the Seebeck coefficient ($S$) for MLG upon doping by BN.
To study the behaviour of doped graphene, we have substituted one and two dimers of boron and nitrogen in our graphene unit cell as shown in Fig. \ref{dg}.
Since boron is an acceptor and nitrogen is a donor, the total number of charge carriers remains unchanged, leading to a gap at the Fermi energy \cite{sm12, RDSM}.
This band gap transforms the metal to a semiconductor, thereby decreasing the electrical conductivity as shown in Fig. \ref{dg} (a) and (c).
The $\frac{1}{\sigma}$ term in Eq. \ref{Smott} increases the Seebeck coefficient. This is evident in Fig. \ref{dg}(b). We can therefore predict that doping graphene with boron and nitrogen will increase the Seebeck coefficient. 

\begin{figure}[!htbp]
\centering \includegraphics[scale=0.3]{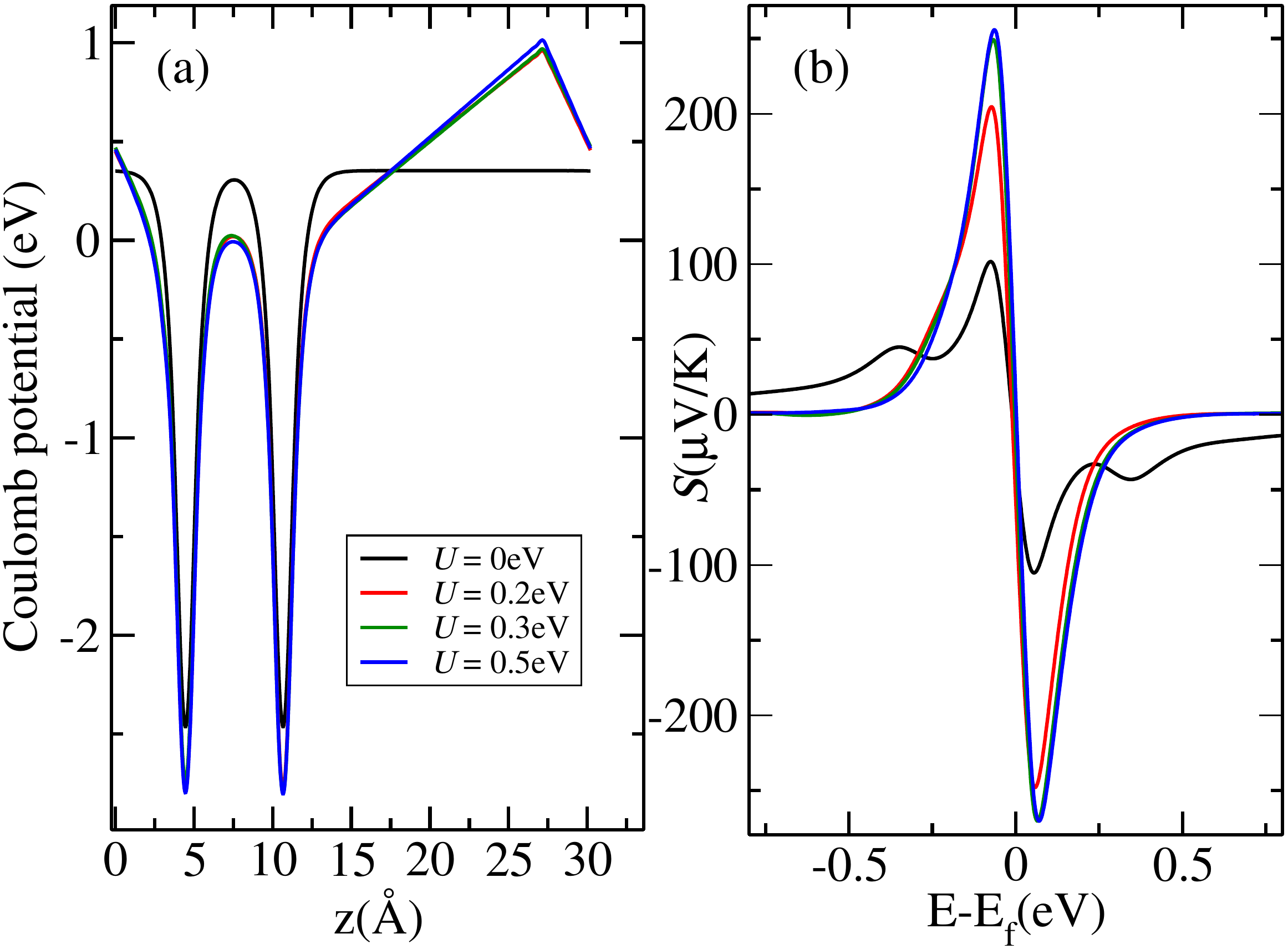}
\caption{\label{blgef} (a) Average Coulomb potential plotted as a function of $z$ for different electric fields $U$, (b) $S$ plotted as a function of energy. 
Differently colored curves refer to different $U$ as in (a).}
\end{figure}

Since a band gap decreases the electrical conductivity thus increasing the Seebeck coefficient,  we apply an electric field perpendicularly to the monolayer graphene
sheets. BLG in an electric field has been shown to have a band gap \cite{hongki07}. We have thus considered the effect of an electric field for three 
different values of external potential ($U$), i.e., $U$=0.2, 0.3, 0.5 eV. The averaged Coulomb potential plotted as a function of its perpendicular length ($z$),  
is shown in Fig. \ref{blgef}(a).
In Fig. \ref{blgef}(b) we plot the Seebeck coefficient of BLG under the influence of an electric field, showing that an increase in the external potential results in an increase in $S$.

Experimentally the effect of electric field on $S$ for BLG has been studied by Wang {\it et al.} \cite{wang11}, showing enhancement of $S$ with increasing 
electric field as our calculations indicate.

\subsection{Impurity scattering in graphene}
Increasing the Seebeck ($S$) coefficient of low-dimensional materials such as graphene has always been a pursuit in thermoelectric applications. Enhancement of $S$ 
by inelastic scattering has been reported recently by Ghahari $et\ al$ \cite{kim2016}. Experimental electrical conductivity studies of graphene 
doped with potassium as a function of charge carrier has been reported by Chen $et\ al$ \cite{chen08}. From Mott's formula Eq. \ref{Smott}, 
it can be readily seen that a decrease in electrical conductivity 
will enhance the Seebeck coefficient. As pointed in the previous section doping of boron nitride decreases the electrical conductivity and hence increases $S$. The unit cells 
used in those calculations are relatively small to mimic the experimental behavior of doping since there is no long-range nature of charge-impurity scattering. 
In order to understand the behavior of electrical conductivity as reported  by Chen $et\ al$ \cite{chen08} one would require a long range nature of impurity scattering and 
hence a large simulation cell making first-principles (DFT) calculations extremely hard. 

As only the $\pi$ states are responsible for transport in MLG and BLG \cite{rdsm-iop2016}, a tight-binding band calculation would be more useful since it would 
allow incorporation
of a very large unit cell to account for the long-range nature of the impurity scattering.
In this section we report the calculation of electrical conductivity of 
graphene with impurities with the help of our tight-binding method using the simple orthogonal nearest-neighbor tight-binding model which has the potential of modeling 
several impurity properties as described by Pedersen $et\ al$ \cite{pedersen13}. 

In this model, we consider an infinite graphene sheet but with one or more atoms replaced by an impurity representing the graphitic impurities. Since we are interested only 
in the transport properties, our Hamiltonian states would consists only of the $p_z$ $\pi$ states. The Hamiltonian would then be written as
\begin{eqnarray}
H = H_0 + H_{imp}
\end{eqnarray}
with $H_0$ defined as
\begin{eqnarray}
H_0 = \sum\limits_{i}\epsilon|i\rangle \langle i| - \sum\limits_{i,j}t_{ij}|i\rangle \langle j| 
\end{eqnarray}
where $\epsilon$ is the on-site energy of carbon, and $t_{ij}$ is the hopping integral which is described by $H_0$ between the $p_z$ $\pi$ states on site $i$ and $j$. 
The added $H_{imp}$ is the impurity Hamiltonian. It depends on which site an atom has been replaced by an impurity and takes the form
\begin{eqnarray}
H_{imp} = \Delta \sum\limits_{i} |i\rangle \langle i |
\end{eqnarray}
where $\Delta$ is the increase or decrease of the on site energy on site $i$. It must be noted that the  difference between $H_{imp}$ and the first term of $H_0$ is that the 
summation in $H_{imp}$ runs over those sites where an atom has been replaced by an impurity whereas in the first term of $H_0$ it runs over all available sites.
\begin{figure}[!htbp]
\centering \includegraphics[scale=0.35]{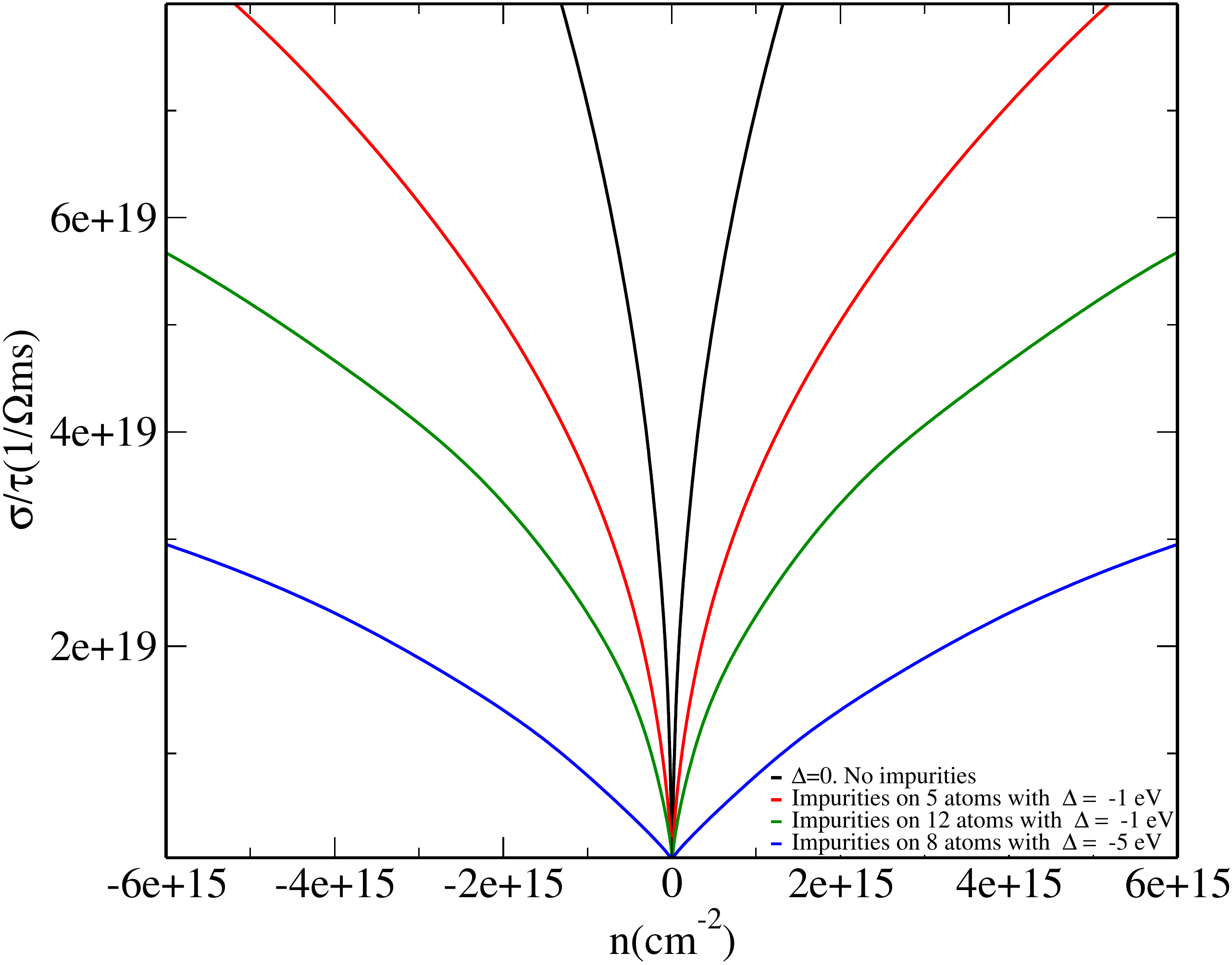}
\caption{\label{sig-imp} Plot of scaled electrical conductivity ($\sigma/\tau$) of graphitic impurities with a supercell of 98 atoms and a K mesh of 150$\times$150$\times$1 with different values of $\Delta$ and different number of atoms replaced with an impurity.}
\end{figure}
From Fig. \ref{sig-imp}  it can be seen that the behavior of electrical conductivity when plotted against charge carrier is that which is observed experimentally 
\cite{chen08} for doped systems. Larger values of $\Delta$ for the same number of atoms replaced by an impurity decreases the electrical conductivity and hence from 
Eq. \ref{Smott} would increase the Seebeck coefficient. One can also see the behavior for electrical conductivity on pristine graphene ($\sigma \sim \sqrt{n}$) 
tending to the linear behaviour of electrical conductivity ($\sigma \sim n$) for graphene sheets with impurities, a fact that has be observed experimentally \cite{kim2}.

\subsection{Mobility ($\mu_{FE}$) of doped and undoped MLG}

\begin{figure}[!htbp]
\centering \includegraphics[scale=0.3]{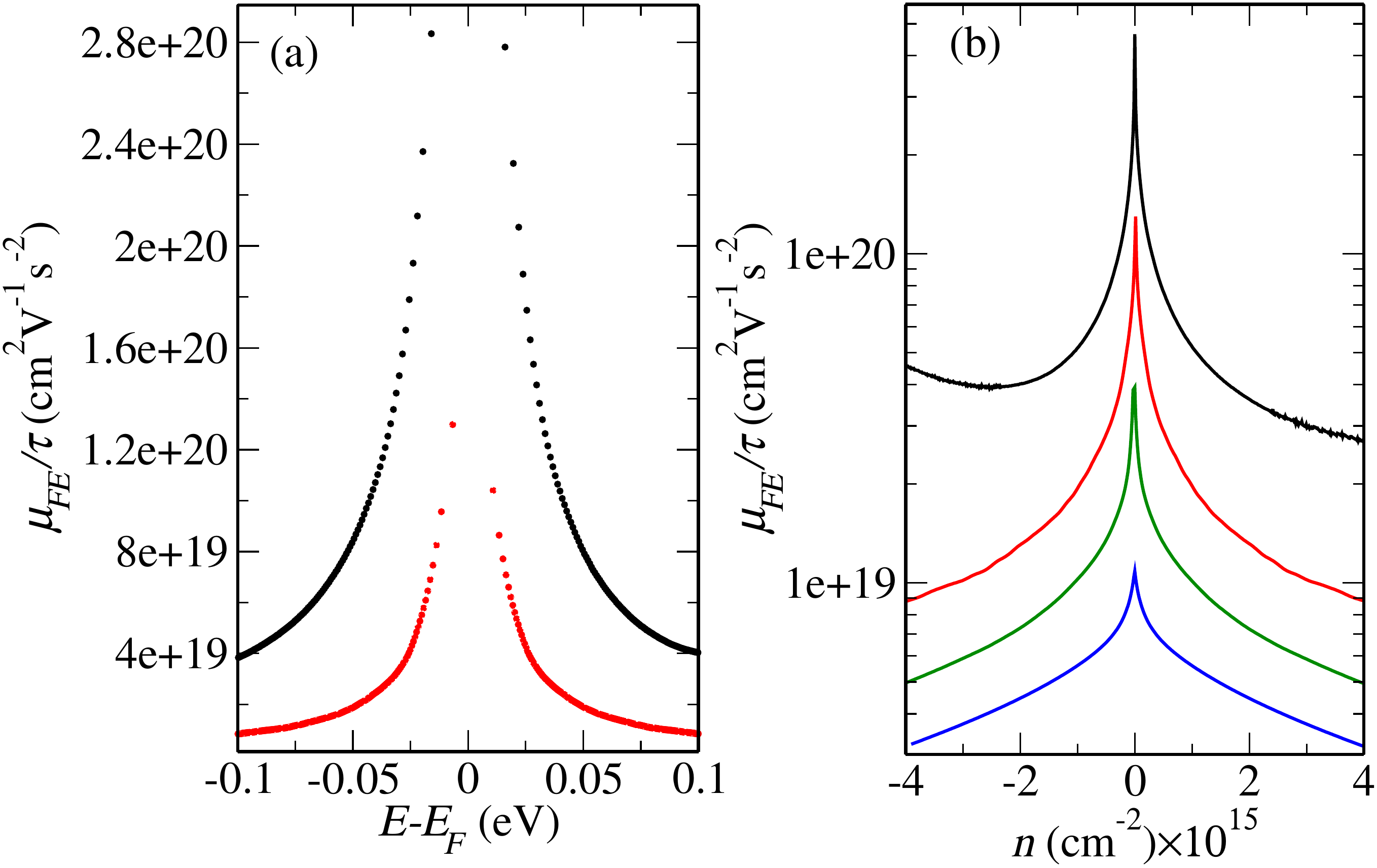}
\caption{\label{mu} (a) Calculated scaled mobility ($\mu_{FE}/\tau$) as a function of energy, and (b) as a function of carrier concentration. The 
black curves are the results for pristine graphene using DFT and Boltzmann transport equations. The red, blue, and green curves refer to results using 
the tight-binding model.
The red curves refer to the results for graphene with impurity, where the onsite energy of one atom in the unit cell is decreased by $1\ $eV, 
whereas the green and blue curves refer to those where the on-site energies of 2 and 4 atoms are decreased by $1\ $eV, respectively.}
\end{figure}

In Fig.\ref{mu}(a,b) we show the scaled mobility ($\mu_{FE}/\tau$) of MLG as function of energy and carrier concentration $n$, calculated from the 
Boltzmann transport equations, showing similar trends as to those seen in experiments by 
tuning the gate voltage and carrier concentration \cite{ponomarenko09,kim2016}, respectively. We have calculated $\mu_{FE}$ using,
\begin{eqnarray}
\mu_{FE}=\frac{1}{e}\frac{d\sigma}{dn}
\end{eqnarray}
It is easy to understand the behavior of $\mu_{EF}$ since the derivative of electrical conductivity with respect to $n$
should be proportional to
$\frac{1}{\sqrt{n}}$. This behavior has been reported by Ponomarenko $et\ al$. \cite{ponomarenko09}.
The method used to introduce impurities is discussed in detail in the previous section. 
Experimental data \cite{tan07,chen08} for graphene samples with increasing doping concentrations have been shown to reduce its mobility by an order of magnitude. 
This effect can be seen in Fig. \ref{mu}(b), where variation of $\mu_{FE}/\tau$ is shown as a function of carrier concentration $n$ plotted in logarithmic scale.
Our results on mobility show similar behavior to that observed experimentally \cite{ponomarenko09,kim2016}.

Expressing mobility in units of  $\tilde{n}=\frac{n}{10^{10}}$ cm$^{-2}$ and $\tilde{\sigma}=\frac{\sigma}{h/e^{2}}$ k$\Omega^{-1}$, and assuming the electron
relaxation time $\tau \sim 1\times 10^{-14}$s, our calculations result in a 
value of $\mu_{FE}\approx 1.6\times10^{4} \frac{\tilde{\sigma}}{\tilde{n}}$ [cm$^2$/Vs] which is close to the earlier estimate of $\mu_{FE}\approx 2.42\times10^{4} 
\frac{\tilde{\sigma}}{\tilde{n}}$ [cm$^2$/Vs] \cite{sdsarma}.

\subsection{Phonon dispersion, Gr\"uneisen parameter, and lattice thermal conductivity}
\subsubsection{Phonon dispersion}

\begin{figure}[!htbp]
\centering \includegraphics[scale=0.3]{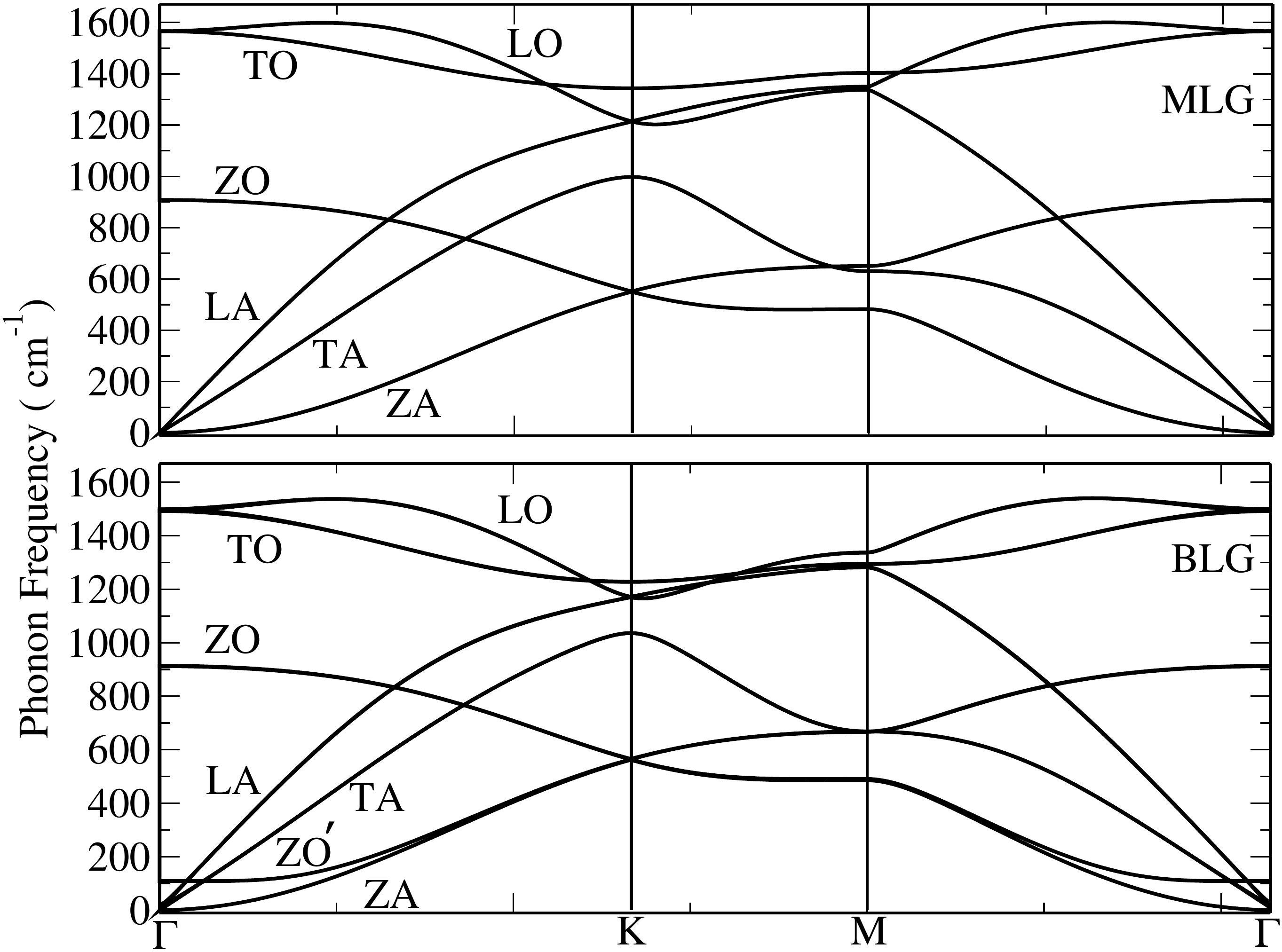}
\caption{\label{pd} Calculated phonon dispersion of MLG (top) and BLG (bottom) along the high-symmetric $q$ points in the hexagonal BZ.}
\end{figure}

In Fig. \ref{pd} we show our calculated phonon band structure along the high-symmetric points in the irreducible hexagonal Brillouin zone (BZ) for the monolayer and
bilayer graphene. 
Accurate calculation of phonon dispersion of MLG and BLG is necessary to understand the thermal conduction in these materials.
Based on the harmonic second-order IFCs, we calculate the phonon dispersion of MLG and BLG along high-symmetric $q$ points obtained within the linear response 
framework by employing density functional perturbation theory (DFPT) \cite{dfpt87}, as implemented in the Quantum Espresso code \cite{giannozzi09} described earlier. 

The out-of-plane (ZA), in-plane longitudinal (LA), and in-plane transverse (TA) modes, which arise from the $\Gamma$ point of the BZ of MLG, correspond to the 
acoustic mode while the remaining branches correspond to the optical modes (ZO, LO and TO) \cite{dejuan2015}. The TA and LA modes show linear $q$ dependence at 
low $q$, as is usually seen 
for acoustic modes. The out-of-plane ZA mode shows a quadratic ($q^2$) dependence, which is a distinctive feature of layered crystals as observed experimentally 
\cite{marzari05, zabel2001}. 
An explanation of this quadratic dependence could be due to the two-dimensional out-of-plane phonon mode and threefold rotational symmetry for BLG 
(sixfold for MLG) \cite{saito98}. 
The LO and TO modes are degenerate  at $\Gamma$ having a frequency of 1580 cm$^{-1}$. Our calculated value of the degenerate frequency is in good agreement with 
the result using inelastic x-ray scattering measurements by Maultzsch $et\ al.$ \cite{maultzsch2004} having a value of 1587 cm$^{-1}$. 
High-voltage transport measurements by Yao $et\ al$ \cite{yao2000} estimated that for graphite the frequency of zone-boundary phonons should be around 
1300 cm$^{-1}$. Our calculations show that at $K$, the BZ corner, the phonon energy of the in-plane transverse optical (TO) mode has a frequency of 1370 cm$^{-1}$ 
for MLG and 1287 cm$^{-1}$ for BLG. This suggests that our calculations agree well with the experiment.
The phonon dispersion of BLG is very similar to that of MLG except for a characteristic feature of an additional low-frequency optical mode with energy nearly 
about 108 cm$^{-1}$ at $\Gamma$. This layer breathing mode arises due to the interlayer movements. 
The phonon dispersions shown in Fig. \ref{pd}, calculated using harmonic IFCs, are consistent with both experimental and previous theoretical  studies 
\cite{yan2008, nika2009, kong2009, yao2000, maultzsch2004}. 

\subsubsection{Gr\"uneisen parameter ($\gamma$)}

\begin{figure}[!htbp]
\centering \includegraphics[scale=0.35]{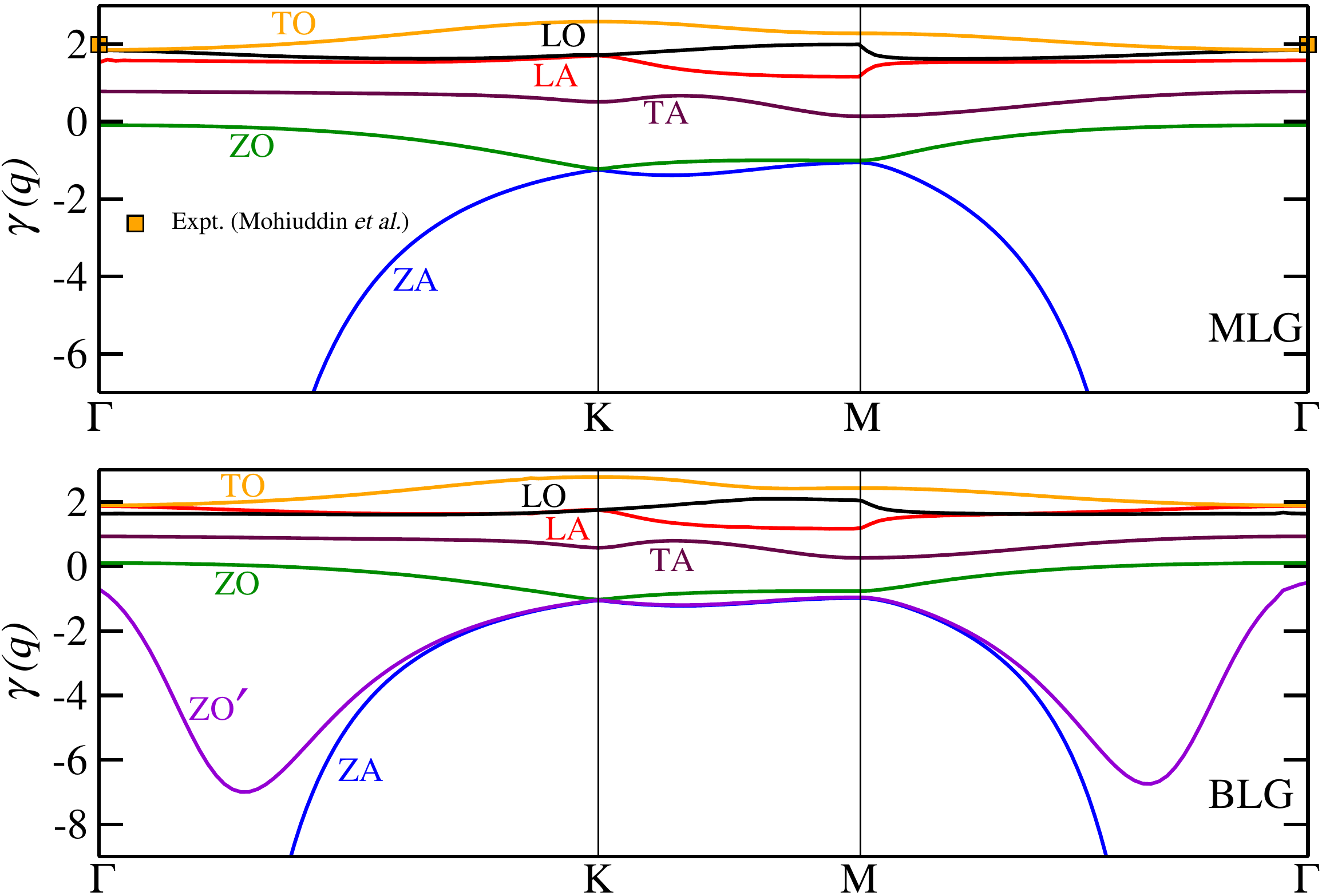}
\caption{\label{gruneisen} Mode-dependent Gr\"uneisen parameters for MLG (top) and BLG (below) along the high-symmetric $q$ points in the first Brillouin zone, calculated from
the first-principles phonon dispersion. The experimental data for MLG corresponding to TO phonons at the $\Gamma$ point are taken from Mohiuddin {\it et al} 
\cite{mohiuddin09}.}
\end{figure}

To carry out a precise calculation of the lattice thermal conductivity, effects from the harmonic and anharmonic 
lattice displacements should be taken into account to include contributions of higher order phonon-phonon scattering processes \cite{broido2005}. 
{Since the Gr\"uniesen parameter ($\gamma$) provides useful information on the phonon relaxation time and the anharmonic interactions between lattice waves 
and the degree of phonon scattering, we have therefore calculated the mode-dependent Gr\"uneisen parameter ($\gamma$) for MLG and BLG.}
 We employ the method as 
developed previously \cite{kong2009,cai14,marzari05} to 
calculate the degree of phonon scattering. It is carried out by dilating the lattice applying a biaxial strain of $\pm$ 0.5\%. In 
Fig. \ref{gruneisen} we show the calculated mode-dependent Gr\"uneisen parameter along the high-symmetric $q$ points. $\gamma$ is expressed as
\begin{eqnarray}\label{gamma}
\begin{split}
\gamma_s(q) & = \frac{-a_0}{2\,\omega_s(q)}\frac{\delta \omega_s(q)}{\delta a} \\
  & \approx \frac{-a_0}{2\,\omega_s(q)}
\Big[\frac{w_+ - w_-}{da}\Big]
\end{split}
\end{eqnarray}
where $a_0$ is the relaxed lattice constant without strain, $\omega$ is the phonon frequency, $\omega_+$ and $\omega_-$ are the phonon frequencies under positive 
and negative biaxial strain, respectively, and $da$ is the difference in the lattice constant when the system is under positive and negative biaxial strain.
MLG and BLG both have negative values for $\gamma$ along the high-symmetric $q$ points for the out-of-plane acoustic (ZA) and optical (ZO and ZO') modes while $\gamma$ has 
only positive values for the in-plane longitudinal and transverse modes. 
Negative (positive) $\gamma$ implies an increase (decrease) in phonon frequency when the lattice constant is increased. 
The slight difference in the Gr\"uneisen parameters in MLG and BLG is, in the case of BLG near the long-wavelength limit ($\Gamma$ point), $\gamma_{ZO}$ 
corresponding to the out-of-plane optical mode changing sign unlike in MLG. This suggests that near the long-wavelength limit the atom vibrations perpendicular to 
the plane of the 
sheet between the two layers lose their coherence hence decreasing the phonon frequencies of $\omega_{ZO}$ when under a biaxial strain.
Since the TO mode in graphene is Raman active at the $\Gamma$ point, the Gr\"uneisen parameter can be measured experimentally using Raman spectroscopy. Mohiuddin 
$et\ al$ \cite{mohiuddin09} have measured $\gamma_{E_{2g}}$ to be 1.99, which is in excellent agreement with our calculated value of $\gamma_{TO}=1.85$ 
($\gamma_{TO}=1.89$) for MLG (BLG).
In Table \ref{gtab} we show the phonon frequencies
and the Gr\"uneisen parameters of MLG and BLG at high-symmetric $q$ points $\Gamma$, K, and M for different vibrational modes.

\begin{table}[!h]
\caption{\label{gtab} Calculated phonon frequency $\omega$ (in cm$^{-1}$) and the Gr\"uneisen parameter $\gamma$ of
MLG and BLG at the high-symmetric $q$ points in the hexagonal BZ for different vibrational modes.}
{\scriptsize \begin{tabular}{lccccccccc}
\hline\hline

System & $q$ & \  & ZA & TA & LA & ZO & TO & LO & ZO' \\
\hline
MLG & & \begin{tabular}{c} $\omega$ \\ $\gamma$ \end{tabular} & \begin{tabular}{c} 0 \\ -100 \end{tabular} & \begin{tabular}{c} 0 \\ 0.779 \end{tabular} & \begin{tabular}{c} 0 \\ 1.848 \end{tabular} & \begin{tabular}{c} 907 \\ -0.086 \end{tabular} & \begin{tabular}{c} 1580 \\ 1.850 \end{tabular} & \begin{tabular}{c} 1580 \\ 1.605 \end{tabular} & \begin{tabular}{c} - \\ - \end{tabular} \\
  & $\Gamma$ &  &   &   &   &   &   &   &   \\
BLG &   & \begin{tabular}{c} $\omega$ \\ $\gamma$ \end{tabular} & \begin{tabular}{c} 0 \\ -50 \end{tabular} & \begin{tabular}{c} 0 \\ 0.936 \end{tabular} & \begin{tabular}{c} 0 \\ 1.640 \end{tabular} & \begin{tabular}{c} 915 \\ 0.110 \end{tabular} & \begin{tabular}{c} 1540 \\ 1.892 \end{tabular} & \begin{tabular}{c} 1544 \\ 1.878 \end{tabular} & \begin{tabular}{c} 108 \\ -0.498 \end{tabular} \\
\hline
MLG & & \begin{tabular}{c} $\omega$ \\ $\gamma$ \end{tabular} & \begin{tabular}{c} 545 \\ -1.245 \end{tabular} & \begin{tabular}{c} 1000 \\ 0.510 \end{tabular} & \begin{tabular}{c} 1230 \\ 1.713 \end{tabular} & \begin{tabular}{c} 545 \\ -1.245 \end{tabular} & \begin{tabular}{c} 1370 \\ 2.584 \end{tabular} & \begin{tabular}{c} 1230 \\ 1.713 \end{tabular} & \begin{tabular}{c} - \\ - \end{tabular} \\
  & K &  &   &   &   &   &   &   &   \\
BLG &   & \begin{tabular}{c} $\omega$ \\ $\gamma$ \end{tabular} & \begin{tabular}{c} 554 \\ -1.06 \end{tabular} & \begin{tabular}{c} 1046 \\ 0.581 \end{tabular} & \begin{tabular}{c} 1205 \\ 1.753 \end{tabular} & \begin{tabular}{c} 557 \\ -1.025 \end{tabular} & \begin{tabular}{c} 1287 \\ 2.779 \end{tabular} & \begin{tabular}{c} 1206 \\ 1.749 \end{tabular} & \begin{tabular}{c} 554 \\ -1.049 \end{tabular} \\
\hline
MLG & & \begin{tabular}{c} $\omega$ \\ $\gamma$ \end{tabular} & \begin{tabular}{c} 578 \\ -1.057 \end{tabular} & \begin{tabular}{c} 631 \\ 0.139 \end{tabular} & \begin{tabular}{c} 1350 \\ 1.953 \end{tabular} & \begin{tabular}{c} 645 \\ -1.004 \end{tabular} & \begin{tabular}{c} 1430 \\ 2.281 \end{tabular} & \begin{tabular}{c} 1372 \\ 1.184 \end{tabular} & \begin{tabular}{c} - \\ - \end{tabular} \\
  & M &  &   &   &   &   &   &   &   \\
BLG &   & \begin{tabular}{c} $\omega$ \\ $\gamma$ \end{tabular} & \begin{tabular}{c} 478 \\ -0.982 \end{tabular} & \begin{tabular}{c} 672 \\ 0.267 \end{tabular} & \begin{tabular}{c} 1327 \\ 2.040 \end{tabular} & \begin{tabular}{c} 660 \\ -0.762 \end{tabular} & \begin{tabular}{c} 1348 \\ 2.432 \end{tabular} & \begin{tabular}{c} 1363 \\ 1.195 \end{tabular} & \begin{tabular}{c} 485 \\ -0.956 \end{tabular} \\

\hline\hline
\end{tabular}}
\end{table}

Our calculated $\omega$ for MLG are in very good agreement with previous calculations \cite{marzari05}, and also $\gamma$ show excellent agreement with previous 
calculations \cite{kong2009, marzari05} and experiment \cite{mohiuddin09}.
Our results for $\gamma$ of BLG are similar to that of graphite 
\cite{marzari05} but differs from BLG results by Kong {\it et al} \cite{kong2009} in the low-$q$ region for the ZO' and ZA modes. The phonon dispersion of graphite 
\cite{marzari05} is very 
similar to that of BLG, so a similar behavior in $\gamma$ can be expected. We feel that the discrepancy in the BLG results by Kong {\it et al} 
\cite{kong2009}
can be attributed to their unstable phonon frequencies for ZO' and ZA modes near $\Gamma$ point, which may not be correct. {It should be noted that the 
thermal conductivity can be calculated with the help of the mode-dependent Gr\"uneisen parameter using the Callaway-Klemens approach \cite{klemens1951, callaway1959, nika2009}.
However, we have not used this approach but a real-space supercell approach as implemented in the ShengBTE method \cite{ShengBTE}.}

\subsubsection{Lattice thermal conductivity ($\kappa_L$)}

In the ShengBTE method \cite{ShengBTE}, the third-order anharmonic interatomic force constants (IFCs) were also taken into account apart from the usual second-order
harmonic IFCs which
produced the phonon dispersion, in the calculations of thermal conductivity ($\kappa_L$). The third-order anharmonic IFCs were calculated using a finite-difference supercell
approach with a set of displaced supercell configurations depending on the size of the system. We have used a $4 \times 4 \times 1 $ supercell for both MLG
and BLG, which generated 72 and 156 configurations for MLG and BLG, respectively. The three-phonon scattering amplitudes are then computed from a set of third-order
derivatives of energy, calculated from these configurations using the Quantum Espresso code \cite{giannozzi09}.

\begin{figure}[!htbp]
\centering \includegraphics[scale=0.3]{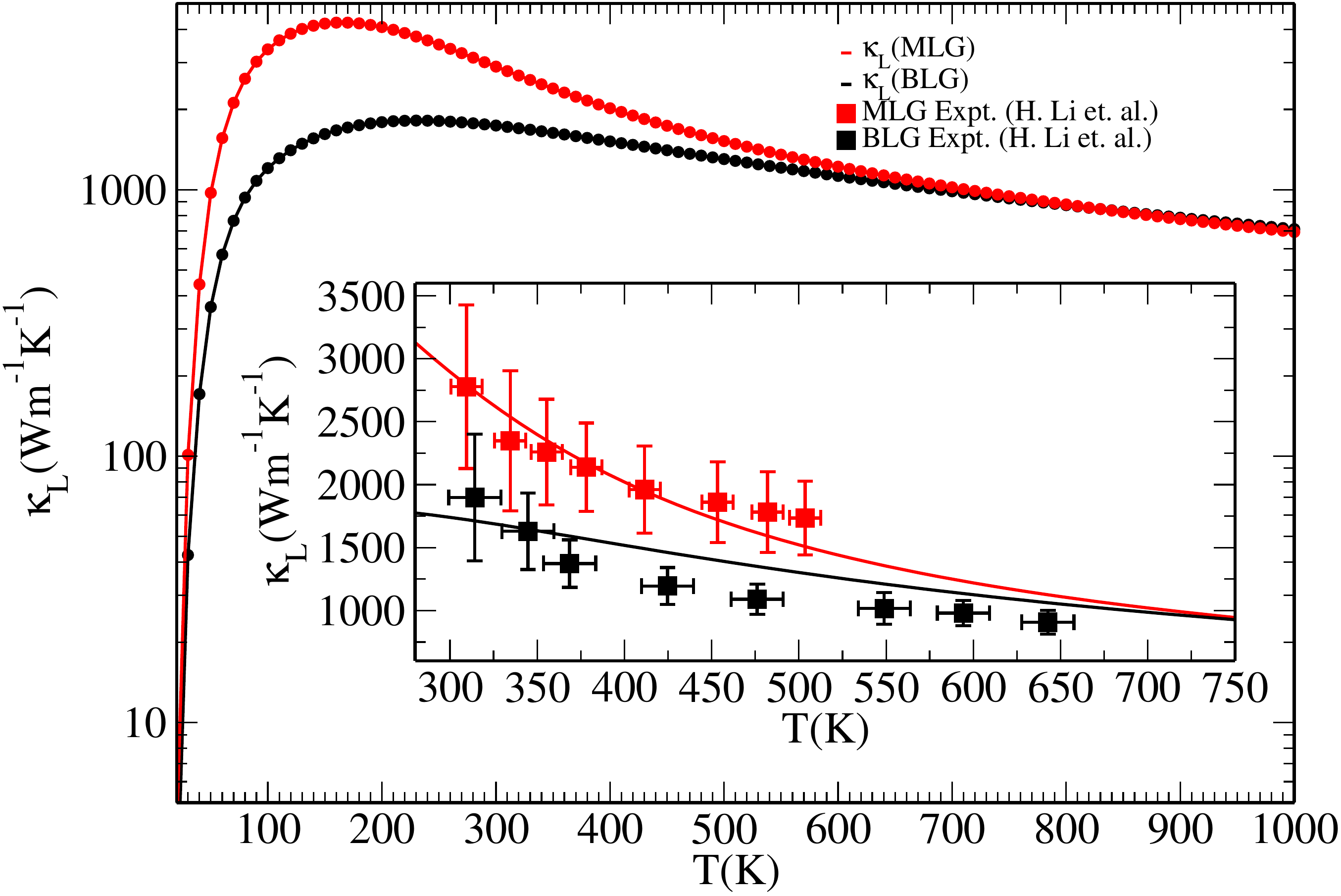}
\caption{\label{k} Calculated lattice thermal conductivity ($\kappa_L$) in log scale of monolayer (red) and bilayer (black) graphene in the temperature range $20\ $K to 
$1000\ $K.
Inset: $\kappa_L$ of MLG and BLG in linear scale in the temperature range $300\ $K to $700\ $K, compared with experimental results \cite{hongyang2014} shown
with red and black square points, respectively.}
\end{figure}

In Fig. \ref{k} we show our calculated lattice thermal conductivity ($\kappa_L$) of MLG and BLG using Eq. \ref{kl} as implemented in the ShengBTE code
\cite{ShengBTE}.
In the inset we compare our results to experimental data of Li {\it et al} \cite{hongyang2014} available in the temperature range 300K to 700K.
Our results are
in very good agreement with experimental measurements.
Graphene at room temperature (RT) has one of the highest know $\kappa_L$.

The experimental results of $\kappa_L$ for MLG \cite{chen2011, chen2012, balandin08, ghosh08, cai10, jauregui10} have shown that for freely suspended samples 
$\kappa_L$  lies between 2000-5000 Wm$^{-1}$K$^{-1}$. This wide variation in experimental estimate of $\kappa_L$ is presumably due to any disorder or residue from fabrication 
leading to an increase in the phonon scattering. We have, therefore, taken the most recent data by Li $et\ al.$ \cite{hongyang2014} to compare with our calculations.
Our calculated $\kappa_L$ at RT of MLG and BLG were found to be 2870 Wm$^{-1}$K$^{-1}$ and 1730 Wm$^{-1}$K$^{-1}$, respectively, which is within the range seen 
experimentally, and are 
in good agreement with the previous literature \cite{nika2009, hongyang2014, pop2012, chen2011, chen2012, ghosh10}. Our calculations also show that at higher temperatures, 
$\kappa_L$ does not change significantly by addition of another layer which is consistent with the report by Koh $et \ al.$ \cite{pop2010}, suggesting 
that $\kappa_L$ between graphene and its environment has a much larger influence than that of individual graphene sheets. 
We find that $\kappa_L$ increases initially from 20 K to 170 K for MLG and to 230 K for BLG, before decreasing. For MLG, if we compare our results to the experimental 
data by Chen $et\ al$ \cite{chen2011}, we find that the maximum values of $\kappa_L$ seen experimentally occuring at a temperature between 150 K and 
200 K are in agreement to our calculations. 
Therefore, calculations involving both harmonic and anharmonic IFCs, solving the BTE for phonons as done in the ShengBTE method, provides an accurate method for the 
calculation of the lattice thermal conductivity.

\section{Summary}
We have reported various transport properties such as electrical conductivity, resistivity, the Seebeck coefficient, mobility and 
lattice thermal conductivity of MLG and BLG graphene using first-principles DFT calculations and Boltzmann transport equations. 
We were able to capture many essential features seen in MLG and BLG, for example $\textendash$ the $\sqrt{n}$ behavior of electrical conductivity and its temperature 
dependence, 
the increase of the Seebeck coefficient with temperature, 
and a linear dependence of the Seebeck coefficient on temperature for a constant chemical potential, as observed in experiments.
For a particular range of chemical potentials we obtained the Bloch-Gr${\rm \ddot{u}}$neisen behavior of resistivity in MLG, where the resistivity increased linearly 
at higher temperatures whereas it showed a $\sim T^4$ behavior at lower temperatures, as observed experimentally. 
We have also observed an order of magnitude decrease in mobility when the energy on impurity sites is decreased, a fact that has been verified 
experimentally.
The Seebeck coefficient was found to increase almost twofold upon doping by boron nitride. 
Our results for graphene with impurities show a systematic decrease in electrical conductivity 
(and hence mobility) when we decrease the on-site terms of particular atoms in the sheet.
We also observe that for a high concentration of impurities, the electrical conductivity was found to change from a $\propto \sqrt{n}$ behavior to a $\propto n$ one.

Our calculated phonon dispersion and Gr\"uneisen parameters with harmonic and anharmonic IFCs, respectively, for both MLG and BLG show good agreement with
available experimental data and previously published calculations. 
We finally show the result of the lattice thermal 
conductivity, calculated using phonon Boltzmann transport theory and first-principles phonon 
bandstructure including both harmonic and anharmonic interactions, showing excellent agreement with recent experimental data \cite{hongyang2014} available in the 
temperature range 300-700 K.
Further experimental measurements are needed to verify the occurrence of a peak in $\kappa_L$
near $T\sim$150-200\ K for both MLG and BLG.

\section{Acknowledgment}
We thank Dr. Jes\'{u}s Carrete for his insightful correspondence on the ShengBTE code for the calculation of thermal conductivity.
All calculations were performed in the High Performance Cluster platform at the S.N. Bose National Centre for Basic Sciences. 
RD acknowledges support through a Senior Research Fellowship of the S.N. Bose National Centre for Basic Sciences. 

\bibliographystyle{apsrev4-1}
\end{document}